\documentstyle [fleqn,12pt,nato] {article}
\newcommand{\cd}{\cdot}

\newcommand{\al}{\alpha}
\renewcommand{\b}{\beta}
\newcommand{\de}{\delta}
\newcommand{\De}{\Delta}
\newcommand{\ep}{\epsilon}
\newcommand{\ga}{\gamma}
\newcommand{\Ga}{\Gamma}

\newcommand{\la}{\lambda}
\newcommand{\Om}{\Omega}

\newcommand{\si}{\sigma}

\newcommand{\th}{\theta}

\newcommand{\ra}{\rightarrow}

\newcommand{\lap}{\triangle}
\newcommand{\arctg}{\mbox{arctg}}

\newcommand{\be}{\begin{equation}}
\newcommand{\ee}{\end{equation}}
\newcommand{\bea}{\begin{eqnarray}}
\newcommand{\eea}{\end{eqnarray}}
\newcommand{\bean}{\begin{eqnarray*}}
\newcommand{\eean}{\end{eqnarray*}}
\newcommand{\dd}{\partial}

\begin{document}

\title{GLOBAL FIELD DYNAMICS AND COSMOLOGICAL STRUCTURE FORMATION}
\author{ Ruth Durrer \footnote{e-mail: durrer@physik.unizh.ch}}.
\affil{ Institut f\"ur Theoretische Physik, Universit\"at Z\"urich,\\
 Winterthurerstr. 190, CH-8057 Z\"urich, SWITZERLAND}

\beginabstract
In this contribution we discuss gravitational effects of global scalar
fields and, especially, of global topological defects.

We first give an introduction to the dynamics of global fields and the
formation of defects. Next we investigate the induced gravitational fields,
 first in a flat background and then in the expanding universe. In flat
space, we explicitly calculate the gravitational fields of exact
global monopole and global texture solutions and discuss the motion of
photons and massive particles in these geometries. We also show that
slowly moving particles and the energy of photons are not affected in
static scalar field configurations with vanishing potential energy.
In expanding space,
we  explore the possibility that global topological defects from a phase
 transition in the very early universe may have seeded inhomogeneities
in the energy distribution which yielded the observed large scale structure
in the Universe,
the sheets of galaxies, clusters, voids ... . We outline numerical
simulations which have been performed to tackle this problem and briefly
discuss their results.
\endabstract

\section{INTRODUCTION}

During this meeting we have learned  about phase transitions,
the formation of topological defects during phase transitions, and the
dynamics  which usually leads to certain scaling laws for the density of
defects and their correlation length. In most of the previous talks
gravitation has been disregarded; it was unimportant for the examples
under consideration. In these lectures we want to discuss the gravitational
interaction of global defects with matter and radiation.

As we shall see, the gravitational coupling strength of topological
defects is of the order of
\be \mu = GT_c^2 = (T_c/m_{pl})^2 ~, \ee
where $T_c$ is the symmetry breaking temperature and $m_{pl}= 1/\sqrt{G}
 \sim 10^{19}GeV \sim 10^{32}K$ is the Planck mass
($\hbar =c=k_{Boltzmann}=1$ throughout). For gravitation to
become important, the symmetry breaking scale thus cannot be
too far below the Planck scale. In the electroweak phase transition, e.g.,
 with $T_c \sim 100GeV$, $\mu \sim 10^{-34}$ and gravity can be ignored
completely.

We shall see later, that topological defects might be responsible for
cosmological structure formation, if they form during a phase transition
at $T_c \sim 10^{16}GeV ~~,~ \mu \sim 10^{-6}$. This energy coincides
roughly with GUT scale. Certainly, this energy scale can never be
probed directly by
accelerators or any present day astrophysical events like supernovae. There
are thus justified doubts if we will ever have a detailed picture of the
physics taking place at these energies. On the other hand, if the ideas
explored here turn out to be correct, and topological defects
due to a phase transition in the very early universe have triggered
cosmological structure formation, then the galaxy distribution in the
universe and the anisotropies in the cosmic microwave background (CMB)
may be relics of  physics at GUT scales!

Since the generation and evolution of defects is quite a generic feature,
we hope that the main results will not depend very sensitively on the
detailed
physical model. In that sense, I think, the scenarios discussed later in
these lectures should be regarded as a kind of toy models which we hope are
capable of capturing the main features, but  we should not expect them
to make predictions / verify observations to much better than within a
factor of two.

In the next section we introduce some generalities on the dynamics of global
fields and defect formation. We mention some important results from
homotopy theory, present the $\si$--model approximation for the dynamics
and discuss Derrick's theorem. In Section~3, we study gravitational effects
of global fields in flat spacetime. We calculate the gravitational
influence of a global monopole and a global texture on matter and radiation.
In Section~4, we investigate global defects in expanding space and
especially the possibility that they may seed the formation of large scale
structure in the universe. We shortly discuss cosmological perturbation
theory, the Harrison Zel'dovich spectrum and numerical simulations of
structure formation. We conclude in Section~5.

\section{ GLOBAL FIELD DYNAMICS AND DEFECT FORMATION}

We consider a scalar field (order parameter) $\phi$ with Lagrangian density
\be
 {\cal L} = {1\over 2}\dd_\mu\phi\cd\dd^\mu\phi - V(\phi) ~, \label{La}
\ee
$\phi \in {\cal V}$, where $\cal V$ is a finite dimensional vector space and
 $ \cd $ denotes a scalar product in $\cal V$. Here "scalar field" does
not refer
to the number of components of the field $\phi$ but to the transformation
of $\phi$ under rotations of physical space: Be $R$ a rotation, then
$ (D(R)\phi)({\bf x},t) = \phi(R^{-1}{\bf x},t)$.

If we quantize the field $\phi$ at finite temperature, we can take into
account the interactions of the $\phi$ particles with the thermal bath by
replacing $V$ by an effective potential $V_T$. The precise form of
$V_T$ depends on $V$ and on the interactions of $\phi$ with other particles,
fermions, gauge bosons ... . In the simplest situation with only the scalar
field $\phi\in {\bf R}^N$ and
\be
  V = {1\over 8}\la(\phi^2-\eta^2)^2  ~ ,
\ee
one finds in one loop approximation the high temperature corrections
\be
  V_T =   {1\over 8}\la(\phi^2-\eta^2)^2
	-  n{\pi^2\over 90}T^4  + {\la(3\phi^2-\eta^2)\over 48}T^2
  +{\cal O}(T)  + V_{(T=0)}
   ~ , \ee
where $n$ denotes the number of helicity states of the $\phi$ field.
At high temperatures, $T^2\gg \eta^2$, the only minimum of $V_T$ is
the field value
$\phi=0$. As the temperature drops below the critical temperature
$T_c = 2\eta$, additional minima at
$\langle\phi\rangle^2 = \eta^2(1-T^2/T_c^2)$ develop and
the vacuum manifold ${\cal N}_T$ (space of minima of $V_T$) becomes an
$(N-1)$--sphere.

More generally, we assume $\cal L$ to be invariant under the action of
some compact Lie group $G$ on $\cal V$, which leaves only $\phi=0$
invariant.
If the vacuum manifold consists only of the invariant element $\phi=0$, the
symmetry is unbroken. Since $V_T$ is temperature dependent, at some other
temperature, the vacuum manifold, ${\cal N}_T$ may become non trivial and
contain an element $\phi_o\neq 0$. Since $V_T$ is invariant under $G$,
 the whole orbit $\{g\phi_o|g\in G\}$ then belongs
to ${\cal N}_T$. If the symmetry group $G$ is maximal, ${\cal N}_T$ consists
just of the orbit of $\phi_o$ which is given by $G/H$, where $H\subset G$
denotes the invariance group of $\phi_o$, $H=\{h\in G| h\phi_o= \phi_o\}$.
The symmetry $G$ is then spontaneously broken to the remaining
symmetry group $H$.

Even though the opposite case can also occur \cite{LP}, we shall
generally assume that the symmetry is restored at high temperatures,
$T>T_c$ and spontaneously broken at lower temperatures $T<T_c$.
If the temperature then falls below the critical temperature, $T_c$, and if
 ${\cal N}_T$ is topologically non trivial, defects of dimension $d$ in
spacetime can form via the Kibble mechanism \cite{Ki}. The collection
of the  different topological defects possible in four spacetime
dimensions is presented in table~\ref{T1}.
\begin{table}[htb]
\begin{tabular}{|l|l|l|l|} \hline
\multicolumn{3}{|c|}{ Homotopy $\pi_n$, ~~~
	dimension in spacetime= d=4-1-n} & appearance \\ \hline
$\pi_o({\cal N}) \neq 0$ & walls form
	& $d=3$ & sheets in space \\
$\cal N$ disconnected &&&\\ \hline
$\pi_1({\cal N}) \neq 0$ $\cal N$ contains
        & strings form
	& $d=2$ & lines in space \\
 non shrinkable circles&&&\\ \hline
$\pi_2({\cal N}) \neq 0$ $\cal N$ contains
	&monopoles form
	& $d=1$ & points in space \\
 non shrinkable 2-spheres&&&\\ \hline
$\pi_3({\cal N}) \neq 0$ $\cal N$ contains
	&textures form
	& $d=0$ & events in spacetime \\
non shrinkable 3-spheres&&&\\ \hline
\end{tabular}
\caption{Topological defects in four dimensional spacetime}
\label{T1}
\end{table}

Since the reader is probably quite familiar with the appearance of domain
walls, strings and monopoles, let me just briefly explain textures:
We consider a field configuration, $\phi$ which is asymptotically
constant (as
it has to be if we require the field to have finite energy).
At fixed time $t$, $\phi$ can then be regarded as map from compactified
space $\bar{\bf R}^3= {\bf R}^3 \cup \{\infty\}$, with
 \[ \phi(\infty) = \lim_{|{\bf x}|\ra\infty}\phi({\bf x})  . \]
Since  $\bar{\bf R}^3$ is topologically equivalent to ${\bf S}^3$, we can
now regard a vacuum configuration, $\phi$, as a map  from ${\bf S}^3$ into
$\cal N$. If the image, $\phi({\bf S}^3)$ (which is of course topologically
again ${\bf S}^3$) is not contractable in $\cal N$, the configuration cannot
evolve into the trivial one, $\phi=$constant, without leaving the vacuum
manifold. Such a configuration is called texture. If $\phi$ has finite
energy, Derricks theorem tells us that it will shrink and eventually
evolve into the trivial configuration,
leaving the vacuum manifold at some spacetime event (with extension of the
order the inverse symmetry breaking scale). This event is the texture
singularity.

We now want to state a few facts from homotopy theory which are commonly
used throughout. Proofs and further information can be found in
 \cite{NS,Jap,Wi}.

In contrast to homology groups, there exists no general algorithm to calculate
homotopy groups $\pi_n$ for $n>1$. Although a lot of research has been carried
out, not even all homotopy groups of the sphere are known!

The following results from homotopy theory of Lie groups are often
useful: Since every compact connected Lie group is a product of some  of the
following groups:
\[ U(n), SO(n), SU(n), Sp(n), Spin(n), G_2, F_4, E_6, E_7 ~~or~~ E_8 ~, \]
it is sufficient to discuss these groups. Here
 $Sp(n)$ denotes the symplectic group of dimension $2n$, $Spin(n)$
denotes the spin group of dimension $n$, i.e., the universal covering
group of $SO(n)$ and $G_2~ ...~ E_8$ are the exception groups. Be $G$ one
of the above groups, then
\[ \pi_1(U(n))= {\bf Z}, ~ \pi_1(SO(n))= {\bf Z}_2 ~ \mbox{ and }
    \pi_1(G)=0 \mbox { for all others. }   \]
\[ \pi_2(G) = 0 ~, \]
\[ \pi_3(G) = {\bf Z} ~,~\mbox{ if } G\neq SO(4) ~\mbox{ and }~
   \pi_3(SO(4)) = {\bf Z} \oplus {\bf Z} ~.\]
\[ \pi_k(U(n)) = 0 \mbox{ for all } k>1 ~.\]
Corresponding identities for direct products follow from
\[ \pi_k(G_1\times G_2) = \pi_k(G_1)\oplus\pi_k(G_2)  ~. \]
The main tool to determine $\pi_n$ are exact sequences.

{\bf Definition:} ~~~ Be $A_n$ a sequence of sets and $\varphi_n$ a mapping
from $A_n$ to $A_{n+1}$. The sequence
\[ ...~ A_n \stackrel{\varphi_n}{\ra} A_{n+1}\stackrel{\varphi_{n+1}}{\ra}
	A_{n+2}~ ... \]
is called exact if
\[ ker(\varphi_{n+1}) = im(\varphi_n) . \]

{\bf Theorem:} ~~~ For a subgroup $H\subset G$ the sequence
\[ ... \pi_n(H)\stackrel{j}{\ra} \pi_n(G) \stackrel{i}{\ra} \pi_n(G/H)
	\stackrel{\dd}{\ra} \pi_{n-1}(H)
	 \stackrel{j}{\ra} \pi_{n-1}(G) ... \]
is exact. Here $j$ and $i$ denote the trivial inclusion map and $\dd$ is
the boundary map.

Since $\pi_2(G)=0$, we thus obtain
\[ \pi_2(G/H) = \pi_1(H) ~~ \mbox{ for all simply connected groups } G; \]
i.e. for all groups with $\pi_1(G)=0$.

\subsection{The $\si$-- model approximation}

If the system under consideration is at a temperature $T$ much below the
critical temperature, $T\ll T_c$, it becomes more and more improbable for
the field $\phi$ to leave the vacuum manifold. $\phi$ will leave the
vacuum manifold only if it would otherwise be forced to gradients of
order $ (\nabla \phi)^2 \sim \la\phi^2\eta^2$, thus only over length scales
of order $l=1/(\sqrt{\la}\eta) \equiv m_\phi^{-1}$ ($l$ is the transversal
extension of the defects). If we are willing to loose the information of
the precise field configuration over these tiny regions (for GUT scale
phase transitions  $l \sim 10^{-30}$cm
as compared to cosmic distances of the order of 1Mpc$\sim 10^{24}$cm !!)
it seems well justified to fix $\phi$ to the vacuum manifold $\cal N$.
Instead of discussing the field equation from (\ref{La}),
\be
	\Box\phi+ {\dd V\over\dd\phi} =0  \label{fe} ~~,
\ee
we require $ \phi \in {\cal N}$. $\cal N \subset V$ is a Riemannian
submanifold with the induced scalar product. The remaining field
equation $\Box\phi=0$ then just demands that
\[ \phi~:~ {\cal M} \ra {\cal N}  \]
is a harmonic map from spacetime $\cal M$ into $\cal N$. There exists
a waste
mathematical literature on harmonic maps and their singularities which
might be useful for us and should be explored \cite{harm}.

The topological defects we are interested in are  singularities of these
maps. When the gradients of $\phi$ become very large, like, e.g., towards
the center of a global monopole, the field leaves the vacuum manifold
and assumes   non vanishing potential energy. If $\phi\in {\cal N}$ is
enforced, a singularity develops by topological reasons.

In the physics literature harmonic maps are known as $\si$--models. They
were originally introduced because of their similarities with non
Abelian gauge theories (the corresponding field equations also contain
non--linear gradient terms). The action of a $\si$--model is given by
\be
 S_\si = \int_{\cal M}g^{\mu\nu}\dd_\mu\phi^A\dd_\nu\phi^Bh_{AB}(\phi)
		\sqrt{|g|}d^4x  ~, \label{Ssi}
\ee
where $h_{AB}$ denotes the metric on $\cal N$ and $g_{\mu\nu}$ is the
metric of spacetime.
Let us consider once more the $O(N)$ example:
\[ V = {1\over 8}\la(\phi^2-\eta^2)^2 ~,~~ \phi\in {\bf R}^N ~.\]
We now fix $\phi$ to lay in the vacuum manifold, ${\bf S}^{N-1}$ with radius
$\eta$, by introducing a Lagrange multiplier.
\[ {\cal L}_\si = \dd_\mu\phi\cd\dd^\mu\phi -\al(\phi^2-\eta^2) ~.\]
Variation w.r.t $\phi$ yields
\be
	\Box\phi +2\al\phi = 0 ~. \label{al}
\ee
We multiply (\ref{al}) with $\phi$ to obtain
$\al= -\phi\cd\Box\phi/(2\eta^2)$.
Inserting this in \ref{al}, we obtain the field equation
\be
   	\Box\phi - {(\phi\cd\Box\phi)\over\eta^2}\phi =0 ~.\label{si}
\ee
In other words, the projection of $\Box\phi$ onto the hyperplane
tangent to the sphere has to vanish, i.e., $\Box\phi=0$ on $\cal N$.
In terms of the dimensionless variable $\b=\phi/\eta$, (\ref{si}) reads
\be
	\Box\b - (\b\cd\Box\b)\b =0 ~,
\ee
which shows that the $\si$--model is scale free.

\subsubsection{ Analytic flat space solutions}
 ~~ \newline
{\bf A global string along the z--axis} is described by the field
configuration $\phi\in{\bf R}^2$:

\bea
\phi(x,y,z) &=& \eta{\bf e}_\rho=\eta(\cos\varphi,\sin\varphi) ,
	 ~~~~ \mbox{$\si$--model}
	\label{string}\\
 \phi(x,y,z) &=& f_S(\rho)\eta(\cos\varphi,\sin\varphi) , ~~~~
	\mbox{full field eqn.}
\eea
where $f_S$ satisfies
\be f''_S +{1\over \rho}f'_S - {1\over\rho^2}f_S +
	{\la\over 2}(f_S^2-1 )f_S = 0 ~, \ee
with boundary conditions $f_S(0)=0$ and $f_S(\infty)=1$.
Here $\rho$ is the cylindrical radius and $\varphi$ the polar angle,
$(x,y)=\rho(\cos\varphi,\sin\varphi)$.
\vspace{3mm}\\
{\bf A spherically symmetric, static global monopole} is described by
the field configuration $\phi\in{\bf R}^3$ with

\bea
\phi(x,y,z) &=& \eta{\bf e}_r=
	\eta(\sin\th\cos\varphi,\sin\th\sin\varphi,\cos\th)
	 , ~~~
	\mbox{$\si$--model}
	\label{monople}\\
\phi(x,y,z) &=& f_M(r)\eta(\sin\th\cos\varphi,\sin\th\sin\varphi,\cos\th)
 , ~~~~ \mbox{full field eqn.}
\eea
where $f_M$ satisfies
\be f''_M +{2\over r}f'_M - {2\over r^2}f_M +
	{\la\over 2}(f_M^2-1 )f_M = 0 ~, \ee
with boundary conditions $f_M(0)=0$ and $f_M(\infty)=1$.
The equations for  $f_S$ and $f_M$ can only be solved numerically.
\vspace{3mm}\\
{\bf A spherically symmetric global texture} is described by the
field configuration
$\phi\in{\bf R}^4$ with
\be
\phi= \eta(\sin\chi\sin\th\cos\varphi,\sin\chi\sin\th\sin\varphi,
	\sin\chi\cos\th,\cos\chi) ~.
\ee
With the ansatz $\chi=\chi(r,t)$, the $\si$--model field equation
(\ref{si}) leads to
\[ (-\dd_t^2 +\dd_r^2 + {2\over r}\dd_r)\chi = {\sin(2\chi)\over r^2} ~.\]
Since the $\si$--model is scale invariant, we further require
 $\chi=\chi(y)$ with $y=t/r$.
In terms of this scaling variable, the equation of motion for $\chi$
reduces to the ordinary  differential equation
\be
  (y^2-1)\chi'' = \sin(2\chi) ~,
\ee
with exact solutions
\[  \chi(y) = 2\arctg(\pm y) \pm n\pi ~. \]
These solutions were originally found by Turok and Spergel \cite{TS}.
To obtain a solution which winds around the three sphere at negative times
and collapses at $t=0$, we patch together $\chi$ as follows:

\be  \chi(y) = \left\{ \begin{array}{ll}
	2\arctg(y) +\pi ~,& -\infty\le y\le 1 \\
	2\arctg(1/y) +\pi ~,& 1\le y\le \infty \\
                  \end{array}\right.    \label{tex}
\ee
The behavior of $\chi$ as function of $r$ for positive and negative times
is shown in Fig.~1. The kink at $r=t$ for positive times is due to the
singularity of the $\si$--model solution at $r=t=0$. It would be softened
in a solution of the full field equations. Physically, this kink represents
the wake of Goldstone bosons in which the massive mode at $r=t=0$ has
decayed and which now travels out with the speed of light.
\begin{figure}
\vspace{7.5cm}
\caption[1]{The function $\chi$ is shown for $t<0$, solid curve, and for
$t>0$, dashed curve.  $\chi$ goes from 0 to $\pi$ for negative times,
i.e. the configuration winds once around ${\bf S}^3$, and from $\pi$
to ${3\over 2}\pi$ and back to $\pi$ for positive times, i.e., no winding.}
\end{figure}
One easily sees that the energy of these three configurations diverges.
For a large ball of radius $R$ one finds
\bea E_{string}(R) =\int T_0^0(z,\rho)={1\over 2}\int(\nabla\phi)^2
	&\propto&  R\log(R\eta) ~, \label{estr}\\
   E_{monopole}(R) =\int T_0^0(r)={1\over 2}\int(\nabla\phi)^2
	& \propto& R ~,  \label{emon}\\
E_{texture}(R) =\int T_0^0(r)=
	{1\over 2}\int[\dot{\phi}^2+(\nabla\phi)^2]
	&\propto& R~~ \mbox{ for } R>>t~. \label{etex} \eea

Before we go on to discuss the gravitational effects of these solutions,
let me briefly note some thoughts concerning Derrick's theorem.

\subsection{Derrick's theorem}

Since it is so simple and beautiful, let me state the theorem with proof.
 \cite{De}\\
{\bf Theorem:}~~~ In $d=3$ dimensions there are no non trivial
static finite energy
solutions for a scalar field whose potential energy is bounded from below.

{\bf Proof:} For static configurations, the variation of the action can be
replaced by the variation of the energy, $E$.
\[ E =  \int [{1\over 2}(\nabla\phi)^2 + V(\phi)]d^3x =I_1+I_2 ~~,\]
with
\[ I_1=\int{1\over 2}(\nabla\phi)^2d^3x ~~,~~ I_2 =\int V(\phi)d^3x~.\]
Without loss of generality, we may assume $V\ge 0$ (otherwise, consider
 $E-V_{\min}$). Then $I_1> 0$ and $I_2\ge 0$. We assume now $\phi(x)$ be a
non trivial solution and consider the scaled configuration
$\phi_\la(x)=\phi(\la^{-1} x)$
For the scaled configuration we have
\[ I_1(\la) \equiv I_1(\phi_\la) = \la I_1 ~\mbox{ and}~~
    I_2(\la) \equiv I_2(\phi_\la) = \la^3 I_2 ~. \]
Therefore
\[ \dd_\la E|_{\la=1} = I_1+3I_2 >0  ~.\]
This contradicts our assumption of $\phi$ beeing a solution. \hfill $\Box$

 From this we can immediately conclude that our solutions for global strings
and monopoles discussed before must have infinite energy.
But also the time dependent texture solution has infinite
energy (\ref{etex}).

Perivolaropoulos\cite{Per} has  put forward the following argument:
In the cosmological context we should truncate the energy at some large
radius $R$, the horizon distance or the distance to the next defect. Then
the variation of the scaled energy yields
\[ \dd_\la E|_{\la=1} = I_1 +3I_2 -R\dd_R(I_1+I_2) ~, \]
which, due to the negative term, can vanish. The second variation shows that
a configuration with vanishing first variation does represent
a minimum of the
truncated energy and thus is stable against shrinking and expansion.

But of course this argument does not explain the existence of the string and
monopole solutions considered previously. Furthermore, the argument
would also
allow for stable static texture solution (with infinite energy).
 There have been some analytical and numerical arguments
\cite{Am,LPr,BCL,PSB,Pe2},
that it is the winding condition that renders the textures unstable. For
winding number $n> 0.5$ textures tend to shrink and for $n< 0.5$ they
tend to expand. Nevertheless, in my opinion, a clear understanding of the
numerical finding that there exist stable static (infinite energy) string
and monopole solutions, but probably no stable static texture solution
is still missing.

\section{GRAVITATIONAL EFFECTS OF SCALAR FIELDS IN FLAT SPACETIME}

\subsection{Generalities}

The energy momentum tensor of a scalar field configuration in the
$\si$--model
approximation is given by
\be
  T_{\mu\nu}^{(\phi)} = \dd_\mu\phi\cd\dd_\nu\phi -
	{1\over 2}g_{\mu\nu}\dd_\la\phi\cd\dd^\la\phi
  \label{emT} ~. \ee

We set
\bea
 \rho = T_0^0 &=& {1\over 2}(\dot\phi^2 + (\nabla\phi)^2)
	\label{rhop}\\
  p  = {1\over 3}T_i^i &=& {1\over 6}(\dot\phi^2 - (\nabla\phi)^2)
	\label{pp}  \\
  \pi_{ij} = T_{ij} -g_{ij}p &=& \dd_i\phi\dd_j\phi -
	{1\over 3}g_{ij}(\nabla\phi)^2
                   \label{pip}  ~. \eea

 For static global field configurations $\rho + 3p =0$. This indicates
that static global field configurations, like an infinite straight string
or a hedgehog monopole, do not gravitationally attract nonrelativistic
particles.

To discuss the gravitational effects of test particles and radiation in
general, we have in principle to solve Einsteins equations,
\be  G_{\mu\nu} = 8\pi GT_{\mu\nu}^{(\phi)} ~,  \label{Ee} \ee
and investigate the geodesics in the resulting geometry. For a typical
field coherence length $l$, we have $ 8\pi GT_{\mu\nu} \sim
8\pi G\eta^2/l^2$.
For a GUT phase transition this is of the order of $10^{-5}/l^2$ ---
 $10^{-4}/l^2$. The induced changes of the metric will thus be small,
of order
$10^{-5}$ --- $10^{-4}$, and we can treat gravity in first order
perturbation theory. I.e., we insert in eqn. (\ref{Ee}) the
unperturbed, flat spacetime,  energy momentum tensor and equate it to the
Einstein tensor
$G_{\mu\nu}$ obtained from first order corrections to the flat metric
(or, in the cosmological context to the Friedmann Robertson Walker metric).

\subsection{Spherically symmetric field configurations}

For the sake of simplicity, we now restrict ourselves to spherically
symmetric
configurations. In first order perturbation theory, the metric can then
 be parametrized by
\be
  g = -(1-2\Psi)dt^2 + (1-2\Phi)\de_{ij}dx^idx^j  \label{mf} ~.
\ee
The linearized Einstein equations yield
\bea
	-\lap\Phi &=& 4\pi G\rho   \label{Phi} \\
        -\lap(\Phi -\Psi) &=& 8\pi G\lap\Pi  ~, \mbox{ where }\label{Psi}
  ~~	\dd_i\dd_j\Pi - {1\over 3}\lap\Pi = \pi_{ij}   ~.
\eea
(In the spherically symmetric case it is always possible to find such
an anisotropy potential $\Pi$.)
For ordinary matter, $\rho \gg \pi_{ij}$, and thus $\Phi = - \Psi$.
$\Psi$ is the relativistic analog to the Newtonian gravitational potential,
and slowly moving matter only couples to $\Psi$. Using the equation of
motion (\ref{si}) for $\phi$, one can show that for static configurations
\be
	\lap\Pi = {1\over 4}(\nabla\phi)^2 = {1\over 2}\rho ~.
\ee
Eqn. (\ref{Phi}) and (\ref{Psi}) then yield $\Psi =0$. This shows again
that {\em nonrelativistic matter is not affected by static global field
configurations}.

It is easy to calculate the connection coefficients (Christoffel symbols)
 from Ansatz
(\ref{mf}). Inserting them into the geodesic equation for a photon moving
with  four velocity
\[  n = (1,{\bf n}) + \de n ~,~~~~   p=En  ~,  \]
one obtains in first order perturbation theory
\bea
  \de n^0 &=& -2\Psi|^f_i  + \int_i^f(\dot{\Phi} -\dot{\Psi})d\la  \label{n0}\\
  \de n_i &=&  \int_i^f\dd_i(\Phi -\Psi)d\la  \label{ni} ~,
\eea
where the integrals are performed along the unperturbed photon trajectory.
The meaning of these quantities is the following: We consider an emitter/
observer of a light ray moving according to the velocity field
\[  u = (1+\Psi, {\bf v}) ~~ \mbox{ with } {\bf v}^2\ll 1  ~. \]
The 0--component of $u$ is determined by the condition $u^2=-1$. The energy
shift of a photon relative to emitter and observer is generally given by
$\de E = (p\cd u)(f) - (p\cd u)(i)$. In our situation this yields
\be \de E/E = \de(u\cd n) = {\bf n\cd v}|^f_i + \Psi|_i^f
	-\int_i^f(\dot{\Phi}-\dot{\Psi})d\la   \label{dE}  ~. \ee
The first term on the right hand side of (\ref{dE}) is the usual, special
relativistic Doppler term. The second term is due to the difference of the
gravitational potential at the position of the emitter and observer, and
the third term is a path dependent contribution due to the change of
the gravitational
potentials during the passage of the photons. Since for static scalar fields
$\Psi=\dot{\Psi}=\dot{\Phi}=0$, the gravitational contributions to $\de E$
vanish in the static situation.

Eqn. (\ref{ni}) is related to light deflection. Be {\bf e} the radial unit
vector. The deflection of a light ray emitted at position $i$ and observed
at $f$ is then given by
\be \al = \de({\bf n\cd e})(i) - \de({\bf n\cd e})(f) =
	-\int_i^f\dd_i(\Phi - \Psi)e_i d\la  ~. \label{ld}
\ee
For the gravitational field from ordinary matter $(\Phi=-\Psi)$, we
recover the old result by Einstein (with the correct factor of 2).

For a slowly moving massive particle  in a weak spherically symmetric
 gravitational field, we make the ansatz
\be u = (1,{\bf 0}) + \de u ~. \ee
 From the geodesic equation, we then obtain
\be \de u^i = -\int\dd_i\Psi dt ~. \label{dv} \ee
Since $\Psi$ vanishes in the static case, we find that slowly moving
particles are not affected by static field configurations.

Taking into account also (\ref{dE}), we thus have proven the following\\
{\bf Theorem:}\hspace{0.2cm} In static scalar field configurations with
negligible potential energy, the gravitational redshift of photons and the
gravitational acceleration of slowly moving particles vanish.

The gravitational field of static configurations thus affects matter and
radiation only by deflection of relativistic particles.

\subsection{Two examples}

To be somewhat more specific, we now want to insert into (\ref{dE}) and
(\ref{ld}) the global monopole and
global texture solutions obtained in the last section.

{\bf Global monopole:}\hspace{0.2cm} From the linearized Einstein equations
we find for the hedgehog monopole solution \cite{d94}
\be \Psi = 0~~,~~~ \Phi = -8\pi G\eta^2\ln(r/l) \equiv -\ep\ln(r/l)
	\label{Gmon}    ~,\ee
where we have set $\ep \equiv 8\pi G\eta^2$ and $l$ is an arbitrary
constant of integration. We consider a light ray passing the monopole
with impact parameter $b$. Its unperturbed trajectory is given by
${\bf x}(\la) = \la{\bf n} + b{\bf e}$. Since the configuration is
static, $\de E$ vanishes. Inserting (\ref{Gmon})
in (\ref{ld})
yields the deflection angle
\be \al_M = -\int \dd_i\Phi{\bf e}^i d\la = \ep\int_{-\infty}^\infty
	{b\over \la^2 +b^2}d\la = \ep\pi  ~.
\ee
This result was originally found by different methods by Barriola and
Vilenkin \cite{BV}.

{\bf Global Texture:}\hspace{0.2cm} For the texture solution (\ref{tex})
we obtain \cite{d90}
\be \Psi = {\ep\over 2}\ln\left( {t^2+ r^2\over t^2}\right) ~~,~~~
    \Phi = - {\ep\over 2}\ln\left( {t^2+ r^2\over l^2}\right) ~.
       \label{Gtex}
\ee
For a light ray passing the texture with impact parameter $b$ at impact
time $\tau$ ($t=\tau+\la$, $r^2=b^2+\la^2$), we find the deflection angle
\bea \al_T &=& -\int_i^f(\Phi-\Psi)_{,i}e^id\la  \nonumber \\
     &\approx& \ep\int_{-\infty}^{\infty}{b\over
  b^2+2\la^2+2\la\tau+\tau^2}d\la   \nonumber  \\
 &=& \ep\pi{b\over\sqrt{2b^2+\tau^2}}~.  \eea
This result was first obtained in \cite{DHJS}.

To calculate the energy shift of a photon passing  the texture, we have to
'renormalize' the result obtained from naively inserting (\ref{Gtex}) in
(\ref{dE}). Due to the unphysical infinite energy of solution (\ref{tex}),
the energy shift contains a divergent logarithmic term which we neglect
in the $\approx$ sign in eqn. (\ref{dET}). (In \cite{d94} this
renormalization is discussed in some detail.)
\be
  {\de E\over E} = \Psi|_i^f + \int_i^f(\dot{\Phi}-\dot{\Psi})d\la \approx
	\ep\pi{\tau\over \sqrt{\tau^2+2b^2}} ~.
	\label{dET} \ee
This result was  first obtained in \cite{TS}.

The interesting difference between the results for monopoles and texture
is due to the time dependence of the latter. This, first of all, yields
a non vanishing energy shift for the texture. An observer receiving
photons from
behind a collapsing texture sees them first redshifted (if they pass the
texture before collapse) and then blueshifted (see Fig.~2). An observer
in perfect alignment with a background quasar and a global monopole sees
the quasar image as Einstein ring with fixed opening angle. In the case
of a global texture, the Einstein ring opens up some time before texture
collapse, reaches a maximum opening angle of the same order of magnitude
as in the monopole case and then shrinks back to a point
\cite{DHJS,d94}.

The gravitational field of our global texture solution (\ref{tex}) also
accelerates slowly moving particles. Inserting (\ref{Gtex}) in (\ref{dv})
 leads to the wellknown result \cite{TS,d90}

\be  {\bf v}(f)- {\bf v}(i) = -\ep\pi{\bf e_r}  ~. \ee
Slowly moving particles around a collapsing texture thus acquire a net
infall velocity of amplitude $\ep\pi$.

\begin{figure}[thb]
\vspace{8.5cm}
\caption[2]{The temperature fluctuation, $\De T/T$, induced by a
spherically symmetric collapsing
texture as function of the impact time of the observer. The solid line
shows the result in expanding space, the dashed line is the flat space
result. The collapse time of the texture is $t_c\approx 20$ (in arbitrary
units). The difference
of the two curves is due to the existence of horizons in expanding space:
Photons, which pass the texture long before of after the collapse are not
influenced in expanding spacetime, but acquire the maximum energy shift in
flat spacetime.}
\end{figure}

\section{GLOBAL DEFECTS AS SEEDS FOR COSMOLOGICAL
STRUCTURE FORMATION}

Many observational results, like Hubble expansion, primordial
nucleosynthesis, the isotropy  and the thermal spectrum of the cosmic
microwave background, confirm the idea that on large
scales the Universe is homogeneous and isotropic. On large scales, the
observable Universe is thus well approximated by a Friedmann universe,
which evolved from a very hot thermal state, the big bang, by adiabatic
expansion.

On smaller scales, clearly, the Universe is lumpy. Laborious mapping of
the 3d galaxy distribution has
 shown that this clumpiness persists on scales up to
(30 -- 50)$h^{-1}$Mpc. The galaxies themselves are arranged in relatively
thin sheets surrounding seemingly empty voids of diameters up to
$50h^{-1}$Mpc.
($1Mpc \approx 3.2\times 10^6 ly \approx 3.1\times 10^{24} cm$)
(see Fig.~3).
\begin{figure}
\vspace{8.5cm}
\caption[3]{The distribution of more than 1000 galaxies of the CfA catalog.
A slice of the universe, about 10'000km/s deep in redshift space and about
$5^o$ thick is shown (from Geller and Huchra~\cite{GH}).}
\end{figure}

With the help of the Cosmic Background Explorer (COBE) satellite,
ani\-so\-tro\-pies have  been found also in the cosmic microwave
background (CMB) which are on the level~\cite{COBE}
\[   \sqrt{\left\langle\left({\De T \over T}\right)^2\right\rangle}(\th)
	\approx 10^{-5}~~, \mbox{ on all angular scales }
	~~ \th>7^o   ~.   \]
These findings support the old idea of Lifshitz \cite{Li} that the cosmic
structures might have formed by gravitational instability from small
initial fluctuations.

Cosmological perturbation theory shows, that perturbations in the
radiation field can not grow substantially. Therefore, $\De T/T$ yields the
amplitude of initial fluctuations $(\de\rho/\rho)_{in} \sim 3\De T/T$.
On the other hand, perturbations in
pressureless matter ($p\ll \rho$, cosmic dust) grow roughly by a factor
$a_0/a_{eq} = z_{eq}+1$, where $a$ denotes the scale factor of the
universe, a subscript $_0$ denotes present time and $_{eq}$ denotes
the time of equal matter and radiation density. If the matter content of the
universe is given by baryons only, $z_{eq} \le 10^3$, and our naive
estimates lead to perturbations which are roughly by a factor 10 too small
to yield the observed structures. However, if we assume that the universe
is dominated by dark matter leading to critical density, $\Om=1$,
we have $z_{eq}\sim 10^4$,
of the correct order of magnitude to lead to the nonlinear clustering
observed today.

There remains one basic ingredient to the gravitational instability
picture:\\
How did the  small initial perturbations of order $10^{-5}$ ---
$10^{-4}$ emerge? Presently two mechanisms
are primarily investigated:
\begin{itemize}
\item Quantum fluctuations 'frozen in' as classical perturbations of the
	energy density  after an epoch of inflation.
\item Topological defects from a phase transition in the early universe.
\end{itemize}
In this workshop, we concentrate on the second possibility. We have seen
in the last section that topological defects yield gravitational perturbations
of the order of $8\pi G\eta^2 \equiv \ep$. To obtain $\ep= 10^{-5}$--$10^{-4}$,
we need a GUT scale phase transition, $\eta\sim 10^{16}GeV$.

\subsection{Scaling}

Let us now assume that on large scales the Universe can be described by
a Friedmann universe with vanishing spatial curvature,
$\Om =1$. The metric of spacetime can then be given by
\be ds^2 = a^2(-dt^2 +\de_{ij}dx^idx^j) ~. \ee
Here $a$ is the cosmic scale factor and $t$ is  conformal time.
It is related to the cosmic time, $t_{cos}$, which has elapsed since the
big bang by
\[ t_{cos}(t)= \int_0^t a(t')dt'   \]
(see also contribution by T.W.B. Kibble).
 To be relevant for structure formation,  topological defects must make up
an approximately constant fraction of order $\ep$ of the total energy
density of the universe. In the cosmological context, we then say that the
defects obey scaling. Let us estimate the energy density of global defects,
neglecting the potential energy:
\[ a^2\langle\rho_{def}\rangle  = {1\over 2}\langle(\nabla\phi)^2 \rangle
	+ {1\over 2}\langle(\dd_t\phi)^2 \rangle  \sim \eta^2/t^2  ~,\]
where we have assumed that $\phi$ changes typically over a horizon scale.
On the other hand, from the Friedmann equation (assuming the scale factor
$a$ to obey a power law), we  have
\[ a^2\rho  = {3\over 8\pi G}(\dot{a}/a)^2 \sim {1\over 8\pi Gt^2} ~,\]
so that  $\langle\rho_{def}\rangle/\rho \sim \ep$.

 From this result one might conclude that all global defects obey scaling.
But the above argument is somewhat too simplistic as the case of global
strings shows: Let us approximate the energy of a global string inside
one horizon volume by the corresponding energy of a  straight
cosmic string in flat space:
\[ E(t) = 2\pi\int_\eta^{t_{cos}} dz r dr(\eta^2/r^2) =
	2\pi\eta^2t_{cos}\log(t_{cos}\eta)~,\]
\[ \mbox{ and thus }~~~ a^2\langle\rho_{str}\rangle
	\sim \eta^2/t^2\log(at\eta)~.\]
In the case of strings, we thus obtain a logarithmic correction term
which, for a GUT scale phase transition, amounts to a
factor of approximately 150 today. For higher $O(N)$ defects like
monopoles, textures and $O(N)$ models with $N>4$, the scaling behavior
becomes clean (see Fig.~4).
\begin{figure}
\vspace{16cm}
\caption[4]{The scaling behavior of $(\rho+3p)a^2$ found numerically in
$(128)^3$ simulations is shown for four different
$O(N)$ models. Time is given in units of the grid spacing $\De x$. For
comparison, the dashed line $\propto 1/t^2$ is shown. For
$N> 3$  scaling is very clean until $t\approx 80$, where finite size
effects can become important.}
\end{figure}

In the case of local defects, only cosmic strings obey scaling. Monopoles
stop interacting soon after formation and then scale like massive particles:
\[ n_M \sim 1/t_c^3(a_c/a)^3 ~,~~~ \rho_M = m_Mn_M \propto a^{-3} ~.\]
The universe at GUT scale is radiation dominated, $\rho \propto a^{-4}$.
Therefore, soon after the phase transition $\rho_M \gg \rho$ leading to
\[ \Om_M(t_0)h^2 \sim 10^{14}(T_c/10^{15}GeV)^3(m_M/10^{16}GeV)
	\gg \Om_0h^2~! \]
This is the famous monopole problem in cosmology \cite{KT}. The reason why
this represents a serious problem is the following:  Imagine some simple,
compact grand unified group $G$, like $SU(5)$, breaking (in one or several
steps) to $H=SU(3)\times SU(2)\times U(1)$. The existence of monopoles at
the end is then determined by the exact sequence
\be \pi_2(G)\ra \pi_2(G/H)\ra \pi_1(H) \ra \pi_1(G) ~. \ee
Since $\pi_2(G)=0$, and for a simple group also $\pi_1(G)=0$ we find
\[ \pi_2(G/H) = \pi_1(H)={\bf Z} ~.\]
Monopoles thus always form. By the analogous sequence for $\pi_1(G/H)$,
\be 0= \pi_1(G)\ra \pi_1(G/H)\ra \pi_0(H) =0 ~, \ee
 we conclude that no strings form. The monopoles are
thus not connected by strings and are stable.

This is a beautiful  example showing that  observations of the present
universe can lead to
 predictions about high energy physics and  cosmology at
GUT scale. The most simple GUT scenario is not compatible with standard
cosmology. One either has to invoke a period of inflation or change the GUT
idea \cite{KT}.

\subsection{Cosmological perturbation theory}

So far, we have only seen that the orders of magnitude  come out
reasonable for structure formation with topological defects from a GUT scale
phase transition. We would like to obtain more precise results. We want to
simulate the evolution of defects, calculate the gravitational fields they
produce, which in turn affect the distribution of matter and radiation. We
want to calculate the induced anisotropies in the cosmic radiation field
and in the matter distribution, $(\De T/T)(t_0,{\bf x,n})$
and $(\de\rho/\rho)(t_0,{\bf x}),~ v_{pec}(t_0,{\bf x})$.

An important tool for this calculation is cosmological perturbation
theory. We do not  develop it here, but just mention the basic
equations which determine our problem. For more details see, e.g.,
\cite{d94}.
\begin{itemize}
\item The equation of motion for the scalar field:
 \bea
   \Box\phi + {\dd V\over\dd\phi} &=& 0 ~~
	\mbox{ (potential model),~~~ or} \\
    \Box\phi -\phi{(\phi\cd\Box\phi)\over \eta^2} &=& 0 ~~
   \mbox{ (sigma model) }. \eea
\item The perturbation of the energy momentum tensor:
  \be
     \de T_{\mu\nu} =T_{\mu\nu}(\phi) + \de T_{\mu\nu}^{matter} ~.
  \ee
\item The linearized Einstein equations:
 \be
     \de T_{\mu\nu} =  \de G_{\mu\nu} ~.
 \ee
\item The equations of motion linearized about the Friedmann background:\\
  - The Liou\-ville equation for photons
 \be
     p^\mu\dd_\mu f + \Ga^i_{\al\beta}p^\al p^\beta{\dd f\over\dd p^i} =0~.
 \label{liouv} \ee
  - The cold dark matter equation of motion, $p=0$,
 \be
  T^{\mu\nu};_\nu=0 ~.
 \label{dm} \ee
\end{itemize}
For a perturbed Planck distribution, the Liouville equation can be cast
into a perturbation equation for the temperature only  \cite{d94,DZ}:
Be $\bf x$ the observer position and $\bf n$ the direction of observation.
If we set $T({\bf x,n}) = \bar{T}(1+m({\bf x,n}))$,
the perturbation equation corresponding to (\ref{liouv}) can be expressed as
\be
  (\dd_t +n^i\dd_i)\chi = -3n^i\dd^jE_{ij} -n^kn^j\ep_{ikl}\dd_lB_{ij}
\label{chi}  ~, \ee
where
\[ \chi = \lap m + \mbox{ (monopole term + dipole term) }, \]
and $E_{ij}$, $B_{ij}$ denote the electric and
magnetic parts of the Weyl tensor.
If we are only interested in the spherical harmonic amplitudes $a_{lm}$ of
$m({\bf n})$  for harmonics higher than the dipole, $l\ge 2$, it is thus
sufficient to determine $\lap^{-1}\chi$. For a fixed observer position a
monopole term can not be distinguised from the background temperature and
a dipole term can be attributed to the  peculiar velocity of the observer.
Therefore, monopole and dipole
terms anyway do not contain information on the temperature fluctuations.

 From (\ref{dm}) we obtain a perturbation equation for the energy density
perturbations, $D$ of the dark matter.
\be
 \ddot{D} +(\dot{a}/a)\dot{D} - 4\pi Ga^2\rho_{dm}D = 4\pi G\dot{\phi}^2 ~.
	\label{D}\ee

 From $\chi$, we can obtain $m=\de T/T$ by inverse Laplacian. The first
information to be compared with observations  are  the power spectra
or, correspondingly, the auto--correlation functions of $\de T/T$
 and $D$. We expand $\de T/T$ in spherical harmonics:
\[ (\de T/T)(t_0,{\bf x,n}) = \sum_{l,m}a_{lm}({\bf x})Y_{lm}({\bf n}) ~.\]
The power spectrum of $\de T/T$ is then given by
\be c_l = {1\over (2l+1)n_x}\sum_{m,x}|a_{lm}({\bf x})|^2 ~,
	\label{cl}\ee
where $n_x$ is the number of observer positions $x$ averaged over, and
$2l+1$ is the number of values $-l\le m\le l$ .
One easily finds  the temperature correlation function \cite{Pa}
\be
\langle (\de T/T)({\bf n})(\de T/T)({\bf n'})\rangle_{{\bf n\cd n'}
	=\cos\th}=
  {1\over 4\pi}\sum_l(2l+1)c_lP_l(\cos\th) ~.
\ee
 $P_l$ denotes the $l$th Legendre polynomial and $\langle \rangle$ indicates
averaging over positions and over all directions ${\bf n,~ n}'$ with
relative angle $\th$.

The power spectrum of  dark matter perturbations (called 'structure
function' in condensed matter physics) is the Fourier transform of the
correlation function. Indicating Fourier transforms by a tilde, we have
\be P(k) \equiv |\tilde{D}(k)|^2 = \tilde{C}(k) ~ \mbox{,  where} \ee
\[ C(r) = \langle D({\bf x})D({\bf x + n}r) \rangle_{\bf x,n} \]
is the correlation function \cite{Ef}.

\subsection{ The Harrison Zel'dovich spectrum}

Let us assume that the only scale in the structure formation problem is
the horizon scale. Then we expect the variance of the mass perturbation on
this scale to be a constant, $A$, independent of time \cite{Ha,Ze}:
\be A =\langle|\de M/M|^2\rangle_{(2\pi/k=t)}
	\approx k^{3}|P(k,t=2\pi/k)| ~.
	\label{M} \ee
Once the perturbations 'enter the horizon', $k>2\pi/t$, their behavior
depends on the expansion law of the background spacetime. From a
simple analysis
of linear perturbation theory one finds that perturbations cannot grow if
spacetime is radiation dominated (M\'ezaros effect, \cite{Me}), and they
grow proportional to the scale factor $a$ if spacetime is matter dominated.
Let us denote by $t_{eq}, a_{eq}$ the conformal time and scale factor of
the universe
at the time when the energy density of  radiation equals that of matter.
During the matter dominated regime the scale factor grows like
$a(t)\propto t^2$. Defining $a_k=a(t=2\pi/k)$, we obtain on scales which
are subhorizon  today ($k>2\pi/t_0$)
\be
  \tilde{D}(k,t_0) \approx \left\{ \begin{array}{lll}
	Ak^{-3/2}(a_0/a_k) =& Ak^{1/2}(t_0/2\pi)^2 &,~ k < 2\pi/t_{eq} \\
	Ak^{-3/2}(a_0/a_{eq}) =& Ak^{-3/2}z_{eq} &,~ k > 2\pi/t_{eq} ~.
  \end{array} \right. \ee
The the Harrison Zel'dovich spectrum can thus be approximated
roughly by the form

\be  P(k) =|D(k)|^2 \approx {Akt_0^2\over 1+ (k/k_{eq})^2} ~ ,\ee
with $k_{eq}=2\pi/t_{eq}$. From large scale structure observations, the
$k^{-3}$ behavior of the spectrum on small scales is approximately
confirmed (the deviations on the smallest scales are probably due
to nonlinear
clustering). The bending of the spectrum on large scales is not yet
observationally confirmed, see Fig.~5, \cite{FDSYH}.
\begin{figure}
\vspace{11.5cm}
\caption[5]{The points are the IRAS redshift space spectrum with $\Om=1$.
The box indicates the power spectrum inferred from the COBE DMR results
with spectral index $n=1$. The solid line is the spectrum of a standard
CDM scenario with $\Om=1$, normalized to the real space variance of IRAS
galaxies, $\si_8 = 0.7$ (this figure is taken from Fisher
et al.~\cite{FDSYH}).}
\end{figure}

Correspondingly, one can show that for a scale invariant spectrum (\ref{M}),
the microwave background fluctuations behave like \cite{Ef}
\be c_l \propto {1\over l(l+1)}~,~~ c_l = {5c_2\over l(l+1)} ~. \ee
Numerical simulations and analytical arguments show that structure
formation by global topological defects leads to an approximately scale
invariant spectrum of perturbations.

\subsection{Numerical Simulations}

\subsubsection{The scalar field:}
The quation of motion of a scalar field in expanding space is given
by
\be
	g^{\mu\nu}\nabla_\mu\nabla_\nu\phi + {\dd V\over \dd\phi} = 0 ~.
  \label{phi}\ee
Defining $\beta = \phi/\eta$ and $m=\sqrt{\la}\eta$, (\ref{phi}) yields for
our $O(N)$ models in a Friedmann universe
\be
  \dd_t^2\beta + 2(\dot{a}/a)\dd_t\beta -\nabla^2\beta =
	{1\over 2}a^2m^2(\beta^2-1)\beta ~. \label{bb}
\ee
This equation as it stands is not tractable numerically in the regime
which is interesting for large scale structure formation. The two scales
in the problem are the horizon scale $t$ and the inverse symmetry
breaking scale, the comoving scale $(am)^{-1}$. At recombination, e.g.,
these scales differ by a factor of about $10^{53}$ and can thus not both be
resolved numerically.

There are two approximations to treat the scalar field numerically. As we
shall see, they are complementary and thus the
fact that both approximations agree with each other within about 10\%
is reassuring.
The first possibility is to replace $(am)^{-1}$ by $w$, the smallest scale
which can be resolved in a given simulation, typically twice the grid
spacing, $w \sim 2\De x$. The time dependence of $(am)^{-1}$ which results
in a steepening of the potential is mimiced by an additional damping term,
$ 2(\dot{a}/a) \ra \ga \dot{a}/a$, with $\ga \sim 3$ \cite{PSR}.
Numerical tests have shown, that this procedure, which usually is
implemented by a modified staggered leap frog scheme \cite{NR}, is not
very sensitive on the values of $\ga$ and $w$ chosen. With this method we
have replaced the growing comoving mass $am$ by the largest mass which
our code can resolve. For a, say $(256)^3$ grid which  simulates
the evolution of the scalar field until today, we have
$256\De x \sim t_0 \sim 4\times10^{17}$sec, so that
$w\sim 2\times 10^{15}$sec,
i.e., $am\sim \eta/z_{rec}\sim 10^{13}GeV$ is replaced by about
$w^{-1}= 10^{-40 }GeV$!

 We believe this mimics the behavior of the field, since the
actual mass of the scalar field is irrelevant as long as it is much larger
than the typical kinetic and gradient energies associated with the field
which are of the order the inverse horizon scale. Therefore, as soon as
the horizon scale is substantially larger than $\De x$, the code should
mimic the true field evolution on scales larger then $w$. But, to my
knowledge, there exists no rigorous mathematical approximation scheme
leading to the above treatment of the
scalar field which would then also yield the optimal choice for $\ga$.

Alternatively, we can treat the scalar field  in the $\si$--model
approximation. This approach is opposite to the one outlined above in
which the scalar field mass is much too small, since the $\si$--model
corresponds to setting the scalar field mass infinity.

The $\si$--model  equation of motion cannot be treated numerically
with a leap frog scheme, since it contains non--linear time
derivatives. In this case, a second order accurate integration scheme
has been developed by varying the discretized action with respect to the
field \cite{PST}.

Initially, the field $\phi$ itself and/or the velocities $\dot{\phi}$ are
 laid down randomly on the grid points. The initial time, $t_{in}$ is
chosen to be the grid size, $t_{in} = \De x$, so that the field at
different grid points should not be correlated. The configuration is
then evolved in time with one of the approximation schemes discussed above.

The two different approaches have been extensively tested, and  good
agreement has been found on scales larger than about 2 -- 3 grid sizes
\cite{BCL1,B}. This is very encouraging, especially since the two treatments
are complementary: In the $\si$--model, we let the scalar field mass $m$ go
to infinity. In the potential approach, we replace it by
$\sim 1/\De x \sim 1/t_{rec} \sim 1/100 ly \sim 10^{-34}$GeV.

The integration of the scalar field equation is numerically the hardest
part of the problem, since it involves
the solution of a nonlinear partial differential equation.

\subsubsection{The gravitational perturbations:}
Once $\beta({\bf x},t)$ is known, we can calculate the energy momentum
tensor
\be T_{\mu\nu}^{(\phi)}= \eta^2[(\dd_\mu\beta\cd\dd_\nu\beta) -
	{1\over 2}g_{\mu\nu}(\dd_\la\beta\cd\dd^\la\beta)] ~. \ee
 From eqn. (\ref{D}) we can further determine the perturbation of the dark
matter energy momentum tensor. The perturbed Einstein equations then yield
an algebraic equation for the
electric part of the Weyl tensor, $E_{ij}$, and an equation of motion for
the magnetic part of the Weyl tensor, $B_{ij}$ (see \cite{DZ}). The
magnetic contributions, which consist of vector and tensor perturbations
only, usually amount to about (10 -- 20) \% of the electric contributions
which are a combination of scalar and vector perturbations of the
gravitational field, see Fig.~6.
\begin{figure}
\vspace{7.5cm}
\caption[6]{The amplitude of the electric and magnetic source terms to
the photon
equation if motion are shown as a function of wavenumber $k$ in arbitrary
scale. For small wavelength (large scales) the magnetic part contributes
about 1/4, decaying to roughly
1/10 on small scales. The quantities graphed are
$\bar{E}= {1\over 3}\sum_i(\partial^jE_{ij})^2 ~,
{}~ \bar{B} ={1\over 5}\sum_{ij}(\epsilon_{jlk}\partial^lB_{ki})^2$.
$\bar{B}$ is represented by the solid line.}
\end{figure}

\subsubsection{The perturbations of the cosmic background radiation
	and the dark matter:}
Using eqn. (\ref{chi}), we finally obtain
\bean {\de T\over T} &=& \lap^{-1}\chi\\
	 &=& \int_{t_i}^t\{n^i\lap^{-1}(\dd^jE_{ij})
  +n^kn^j\lap^{-1}(\ep_{ikl}\dd^lB_{ij})\}(t',{\bf x -n}(t-t'),{\bf n})dt'
  ~. \eean
 From eqn. (\ref{D}), we can calculate  $D$ and the power spectrum
$P(k)=|D(k)|^2$. Patching
together simulations with different physical grid size, we can enlarge the
range of comoving wave numbers $k$ covered, see Fig.~7.
\begin{figure}
\vspace{7.5cm}
\caption[7]{The final dark matter spectrum of density fluctuations for
3 texture simulations with different physical grid sizes patched together.}
\end{figure}

By the decomposition into spherical harmonics,
\be a_{lm}({\bf x}) = \int_{\bf S^2} (\de T/T)(t_0, {\bf x,\bf n})
	Y_{lm}({\bf n}) d\Om ~, \ee
 and (\ref{cl}), we determine the $c_l$'s. Since the spectrum is close to
scale invariant, it is entirely determined by the quadrupole moment, $Q$,
\be Q ={4\pi\over 5}c_2 \approx \al\ep =
	Q_{COBE}=(0.6\pm0.1)10^{-5} ~. \ee
The value of $\al$ above can be obtained by numerical simulations
 \cite{BR,DHZ,PST,DZ} and is typically of the order $0.1\le\al\le 1$.
For textures one finds \cite{DZ}
\be \ep = 8\pi G\eta^2 = (2.2\pm 1)\times 10^{-5} ~. \ee
The shape of the dark matter spectrum is again approximately determined
by scale invariance. The integral of the dark matter perturbation spectrum
over scales larger than $R$, determines the mass variation, $\si^2(R)$
over these scales. The comparison of this dark matter mass variation with
the observed  variation of the galaxy distribution yields a scale
dependent bias factor, $b(R)$.
The bias factors obtained this way are of the order $b \sim 2$ -- 4, which
is somewhat larger than expected \cite{CO}. The global defect models
normalized to the COBE results for the microwave background fluctuations
probably yield somewhat too small perturbations in the dark matter.
Nevertheless, the uncertainties concerning the bias factor and the
nonlinear physics going into the calculation of the bias factor seem to me
to leave room for doubts. It would be more convincing to rule out the
the scenarios from the completely linear determination of the microwave
background fluctuations alone. So far, only the gravitational interaction
of the radiation field with perturbations has been taken into account.
To calculate
$\de T/T(\th)$ on scales, $\th < 2^o$, which enter the horizon before
recombination, when baryons and photons still are a tightly coupled fluid,
the baryon photon interaction has to be taken into account and the
recombination process has to be modeled. For pure CDM without scalar field
this calculation has been performed on different levels of accuracy
\cite{SG,HS,uros}. For global defect induced fluctuations, intermediate
and small scale anisotropies have only been approximated in the case when
the universe is reionized at some early redshift, $z>100$, and  baryons and
photons are coupled again via Thompson scattering. In this case, photon
diffusion severely damps fluctuations on scales smaller than the horizon
at $z\sim 100$, i.e., on all scales smaller than about $5^o$
\cite{d93,d94,CFGT}.   One of the missing pieces in the global defect
scenarios is thus a detailed calculation of the microwave background
spectrum on angular scales $\th < 2^o$ or $l>100$ for a non reionized
universe.

On the other hand, the CMB fluctuations, are not determined by the
spectrum $c_l$ alone. The spectrum just yields the two point correlation
function which determines the fluctuations only if they are Gaussian
distributed. In general the $a_{lm}$ also yield  non zero higher
correlation functions. The skewness $S$ and the kurtosis, $K$
of the distribution of $2^o \times 2^o$ pixels are
found to be \cite{DHZ}
\bean  S &=& -4 \pm 2.3  \\
	K &=& 32 \pm 29  ~.	\eean
 The deviation from Gaussian distribution is also shown
in Fig.~8. These higher order correlations are an important mean to
distinguish models with global defects from models
with initial fluctuations from an inflationary epoch which usually yield
Gaussian fluctuations.
\begin{figure}
\vspace{8.5cm}
\caption[8]{The pixel distribution of $\De T/T$ in a synthesized map of
microwave background fluctuations for the texture scenario of structure
formation. The dashed line shows a Gaussian with the same width and the
same number of pixels. The negative skewness and the positive kurtosis are
clearly visible.}
\end{figure}

\section{CONCLUSIONS}

We have discussed gravitational interaction of global scalar fields with
matter and radiation. We have found that static global field configurations
with vanishing potential energy do not affect slowly moving particles
and do not redshift photons.

Topological defects which form during phase transitions
in the early universe can have important cosmological consequences. For
gravitational interactions of the defects with the cosmic matter and
radiation
to be relevant, the defects must form due to a phase transition at GUT
scale. In this case they may even seed the formation of  cosmological
large scale structure. Even though it is not yet clear if defect induced
structure formation scenarios do work out in detail, up today they remain
an intriguing alternative to initial fluctuations from inflation since
they also yield a scale invariant spectrum of perturbations.
\vspace{1cm} \\
{\large \bf Acknowledgement:}\hspace{1cm} I'm grateful to my collaborator
Zhi--Hong Zhou, who helped me in preparing some of the figures for these
proceedings. I have learned a lot from discussions  with many of the
participants, especially Robert
Brandenberger, Pedro Ferreira, Tom Kibble, Andrew Little, Leandros
Perivolaropoulos, Tomislav Prokopec,
Paul Shellard and Neil Turok. Finally, I  want to express special thanks
to  Ann Davis, the organizer of this lively and stimulating meeting.

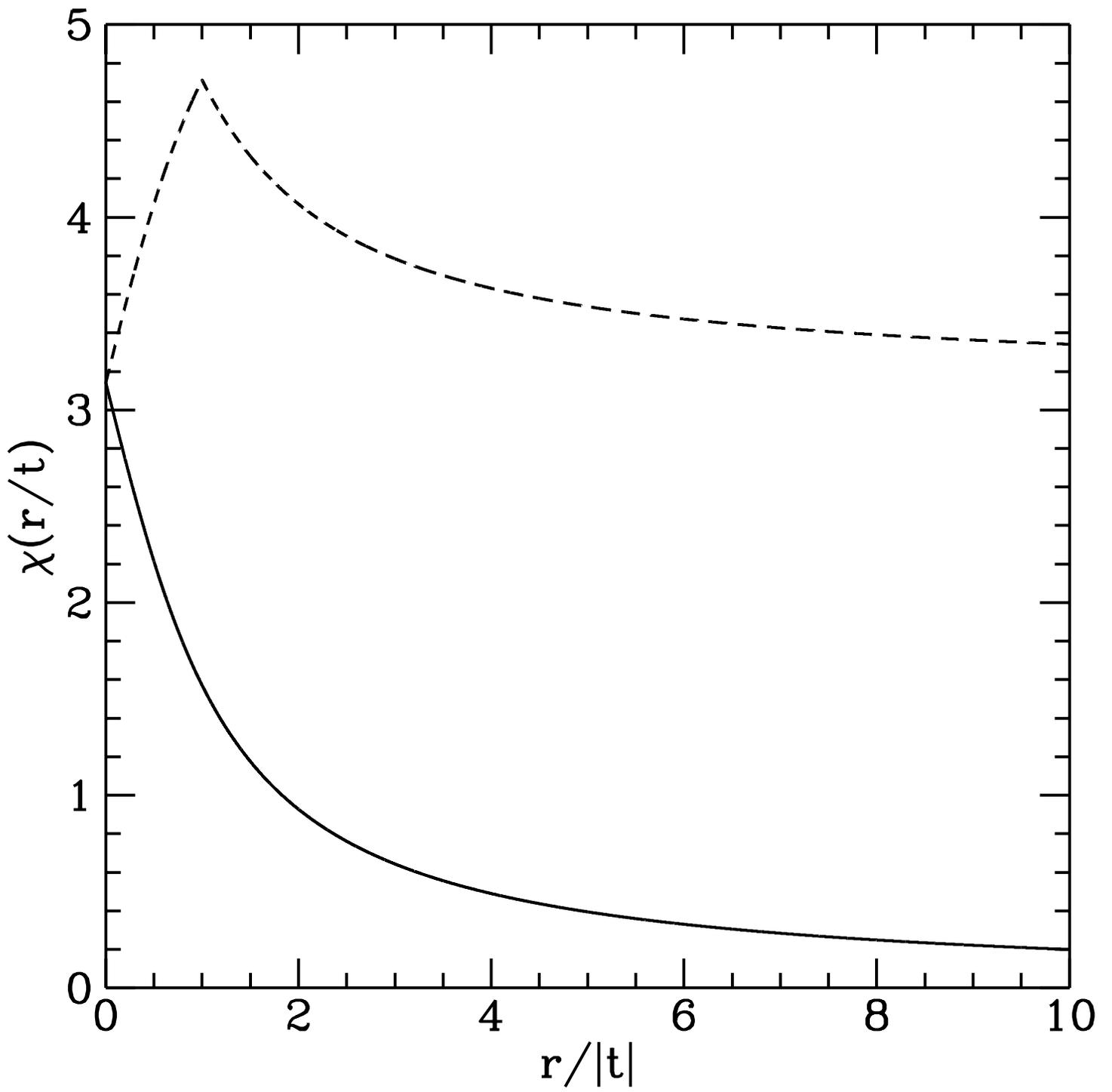
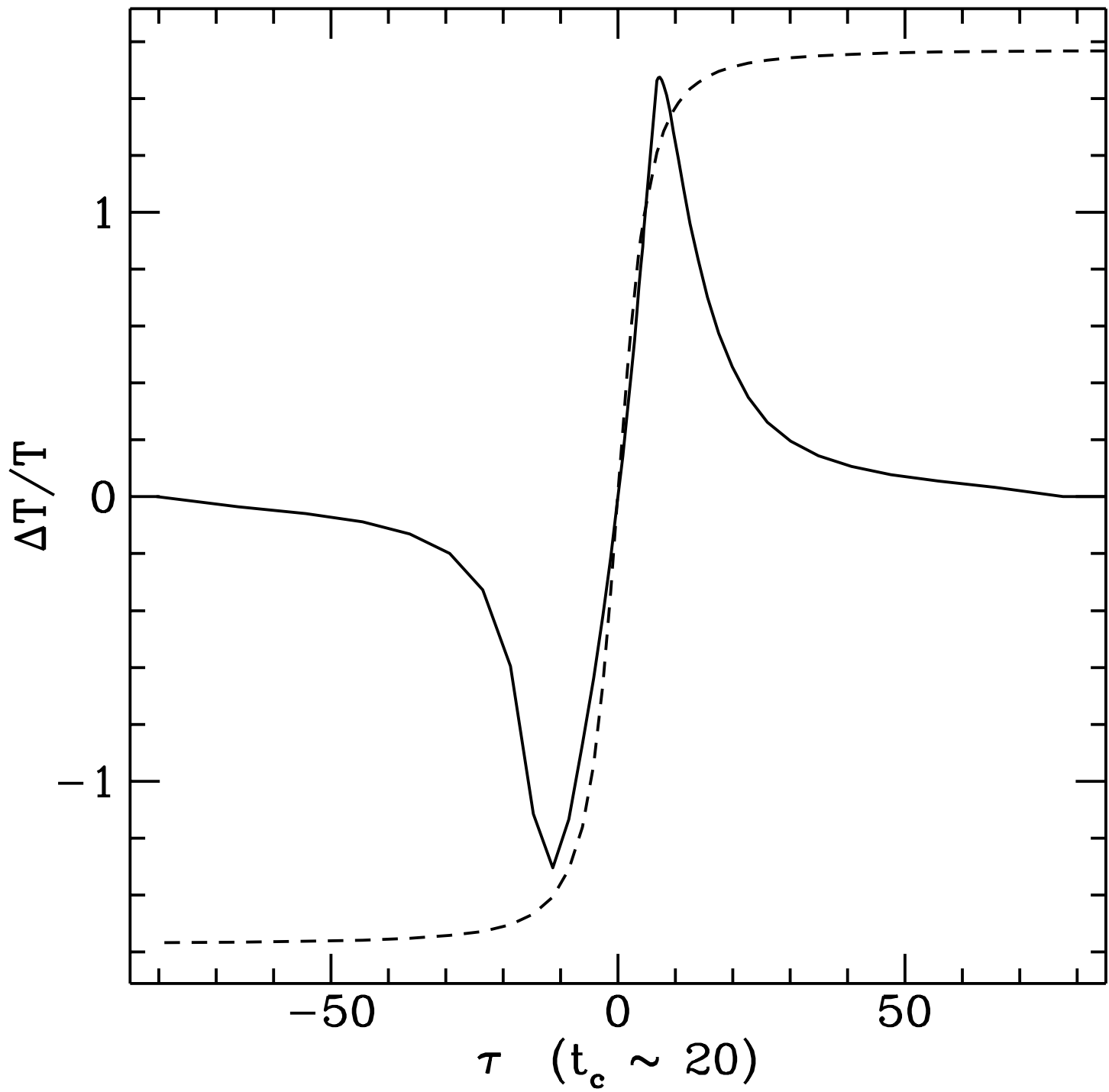
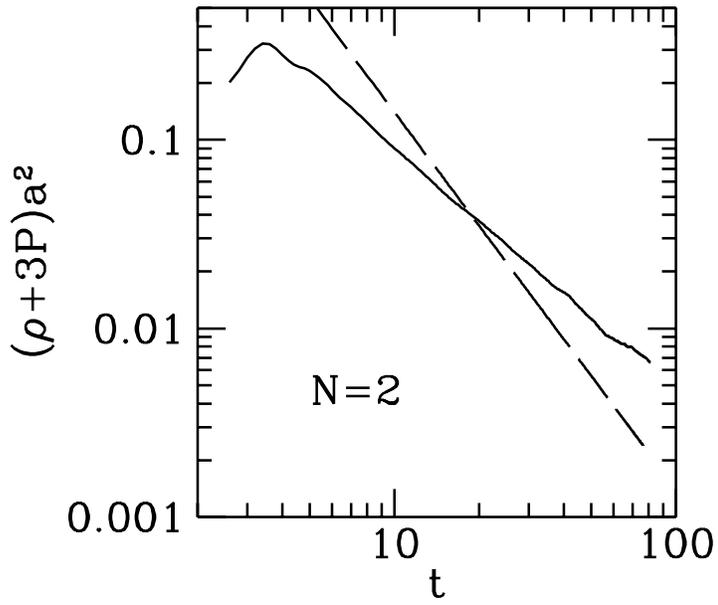
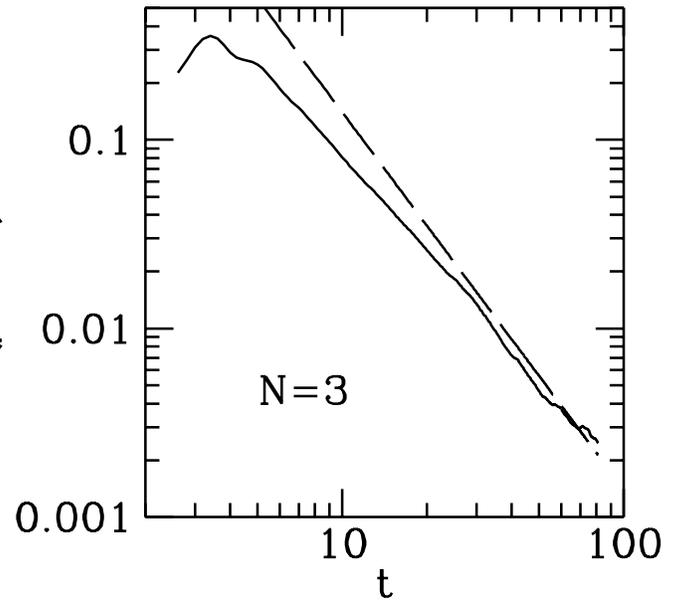
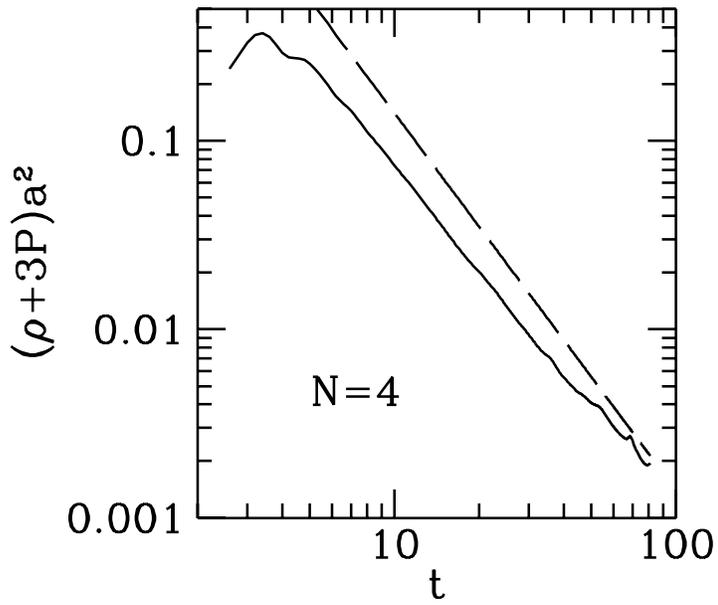
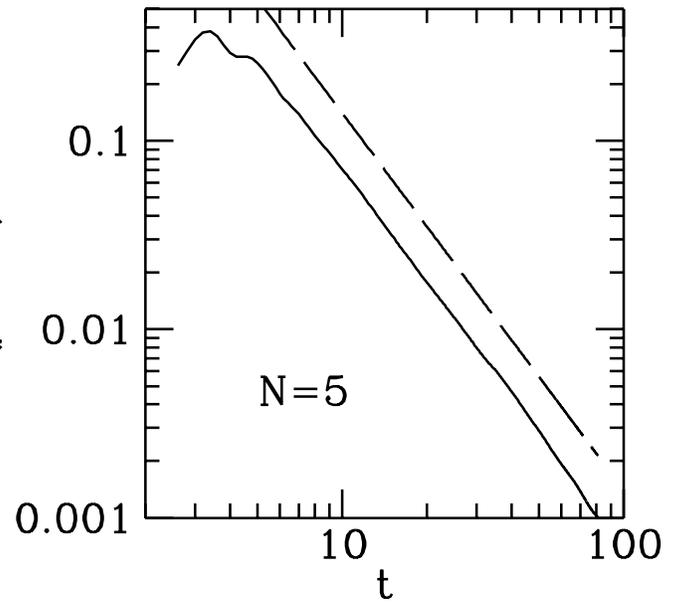
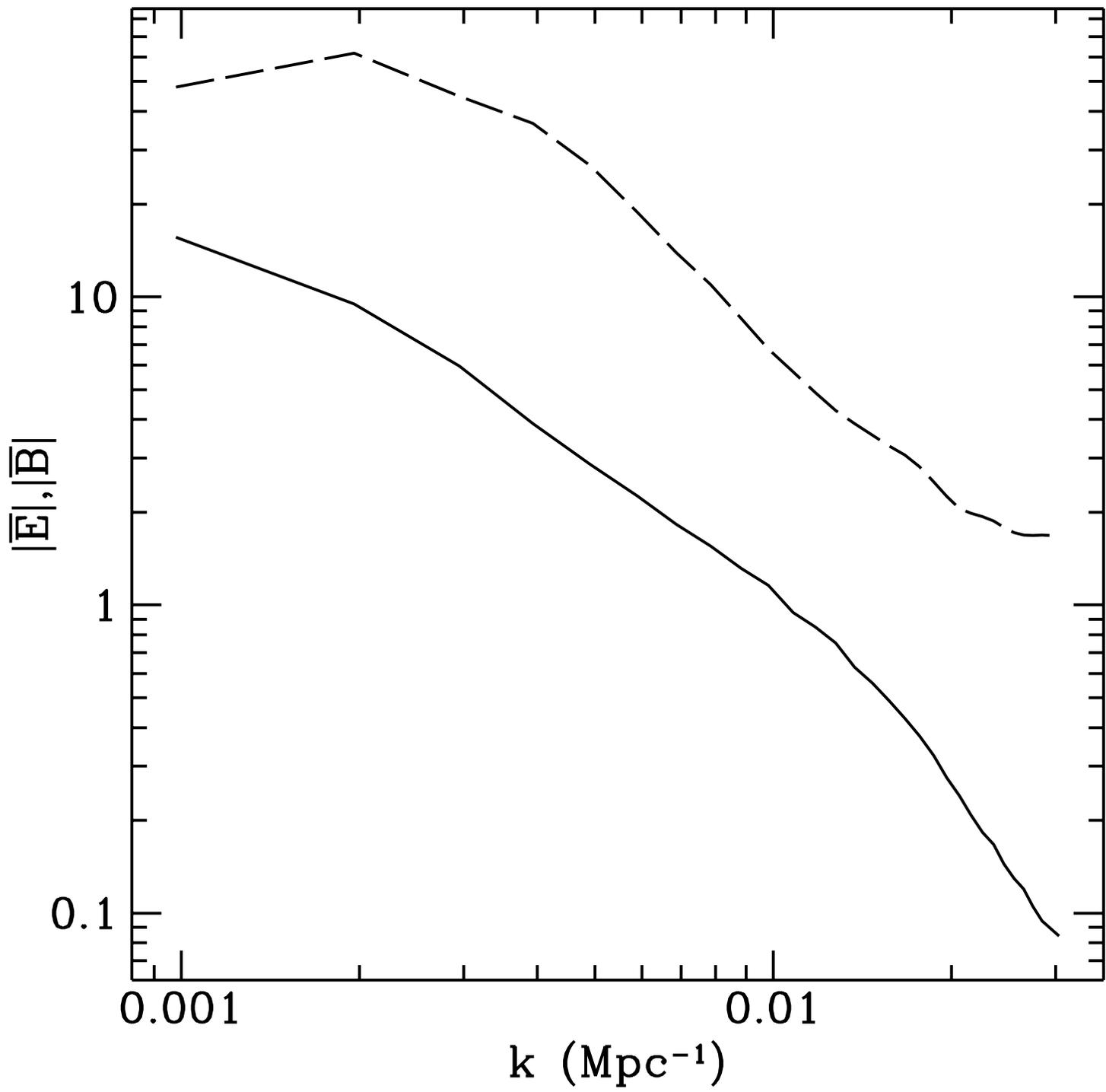
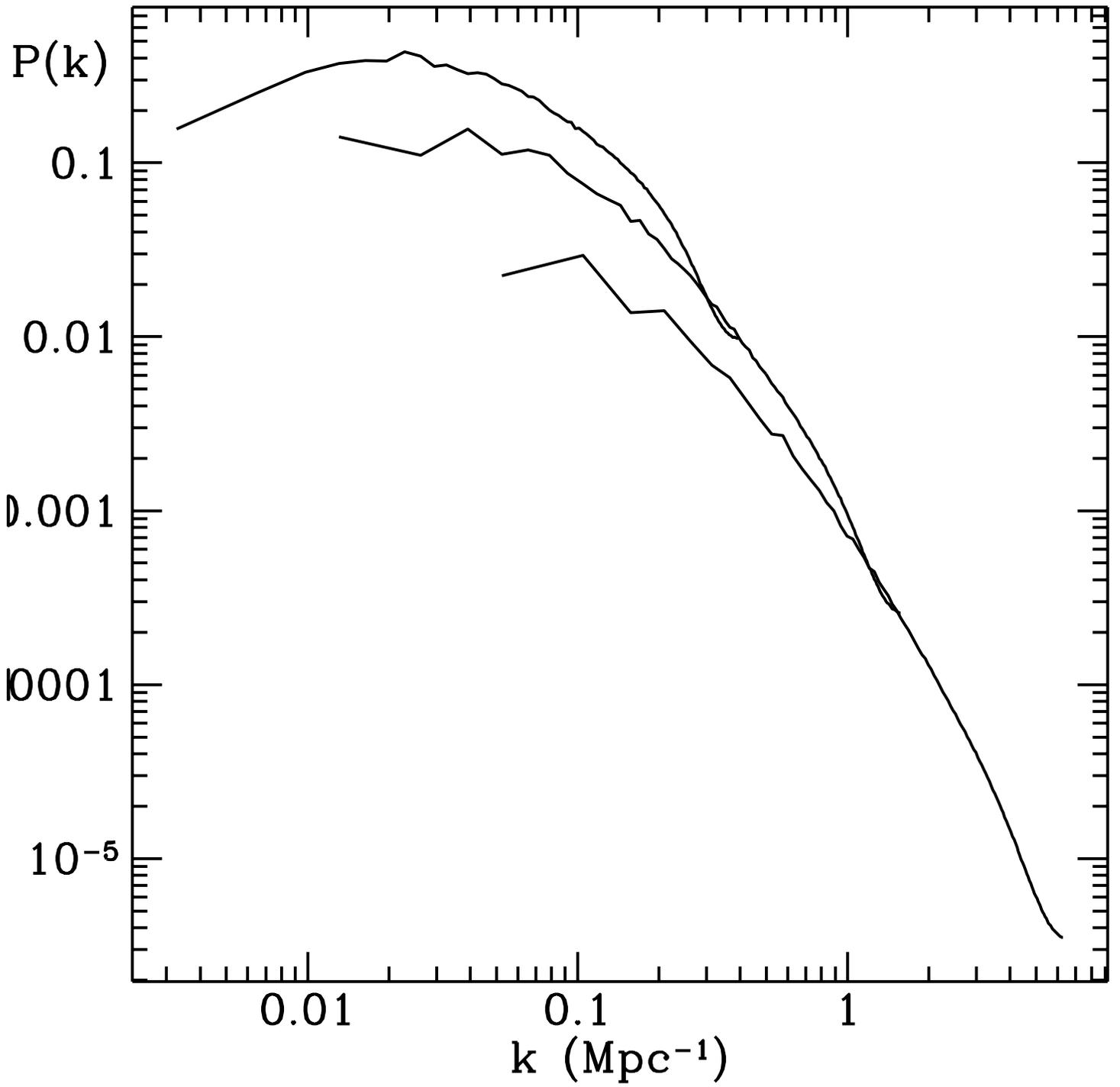
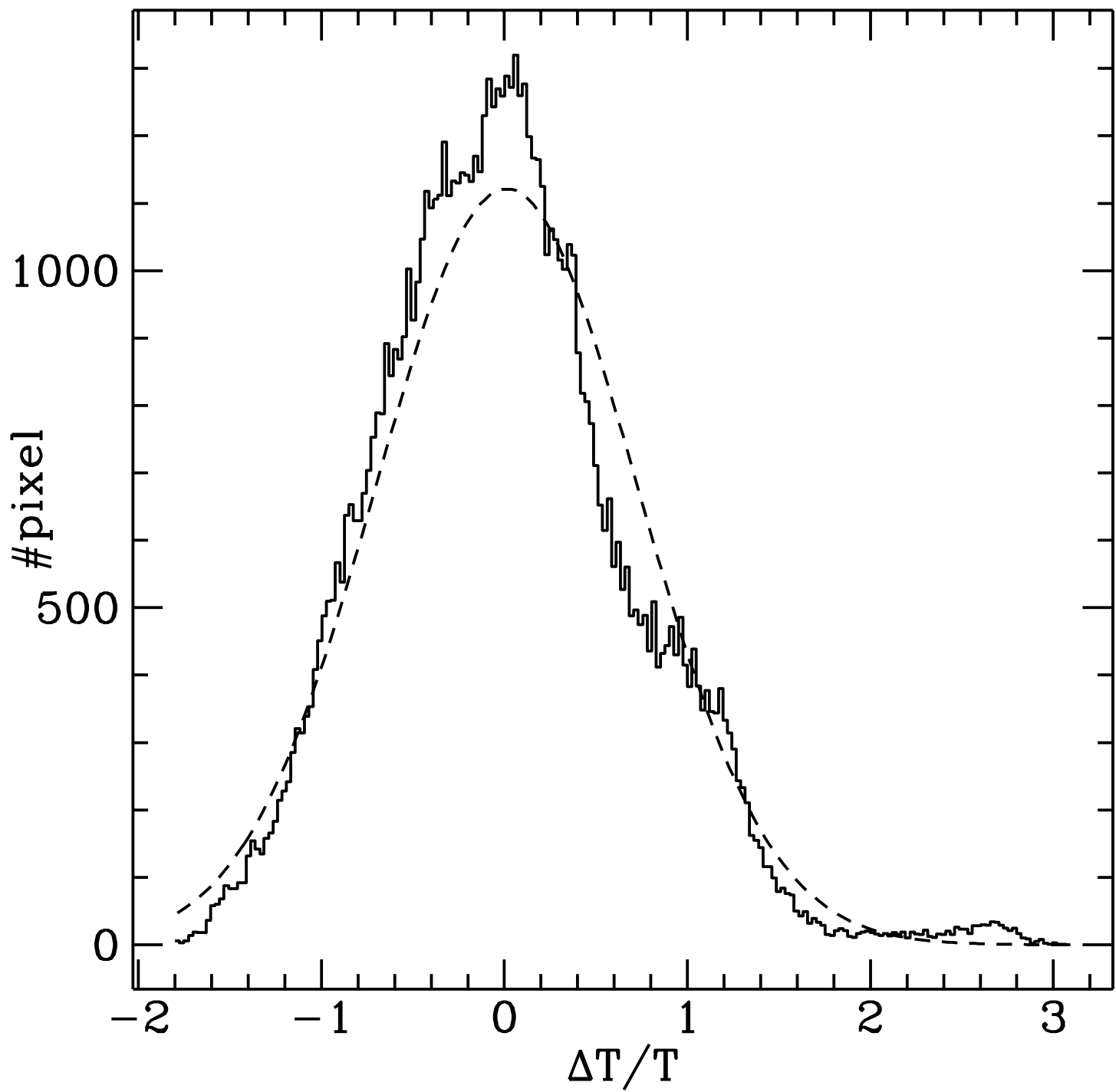
\end{document}